\documentclass[10pt]{iopart}
\usepackage{graphicx}
\usepackage{multirow}

\usepackage{bm}

\newcommand{\abs}[1]{\vert #1\vert}

\begin{document}

\title{Composition and Stacking Dependent Topology in Bilayers from the Graphene Family}

\author{Adrian Popescu$^{1}$, Pablo Rodriguez-Lopez$^{2,3}$, and Lilia M. Woods$^{1}$}
\address{$^{1}$Department of Physics, University of South Florida, Tampa, Florida 33620, USA}

\address{$^{2}$Materials Science Factory, Instituto de Ciencia de Materiales de Madrid (ICMM),
Consejo Superior de Investigaciones Científicas (CSIC),
Sor Juana Inés de la Cruz 3, 28049 Madrid Spain.}
\address{$^{3}$GISC-Grupo Interdisciplinar de Sistemas Complejos, 28040 Madrid, Spain}

\begin{abstract}
We present a compositional and structural investigation of silicene, germanene, and stanene bilayers from first-principles. Due to the staggering of the individual layers, several stacking patterns are possible, most of which are not available to the bilayer graphene. This structural variety, in conjunction with the presence of the spin-orbit coupling, unveil a diversity of the electronic properties, with the appearance of distinct band features, including orbital hybridization and band inversion. We show that for particular cases, the intrinsic spin Hall response exhibits signatures of non-trivial electronic band topology, making these structures promising candidates to probe Dirac-like physics.
\end{abstract}


\maketitle

\ioptwocol

\section{Introduction}

The discovery of graphene has inspired the pursuit of other 2D systems with honeycomb atomic arrangements which may host novel Dirac-like properties. Similar to C, the first element in group-IV from the periodic table, Si, Ge, and Sn are also found to exist as 2D materials with hexagonal lattice structure giving rise to silicene, germanene, and stanene, respectively. Growing experimental work on synthesis and characterization of these 2D structures on various substrates \cite{LeFay2012,Takamura2012,Kawai2012,silicenereview,LeLay2014,Pirri2015,Jia2015,germanenereview,Shukla2016} suggests that the graphene family is expanding. Comparing the different graphene-like materials one finds that silicene, germanene, and stanene have distinct from graphene features. Specifically, the C atoms take a planar arrangement in graphene, while Si, Ge, and Sn form a staggered 2D hexagonal structure \cite{Ciraci2009,Bechstedt2013,Smallreview2015}. Another distinction is the significant spin orbit coupling (SOC) for silicene, germanene, and stanene as compared to the vanishingly small SOC of graphene. 

Recent reports have shown that these new materials exhibit quantum Hall effect phases resulting in rich physics in their electronic and optical response properties, Casimir interactions, and superconductivity \cite{Ezawa2012,Ezawa2013,Nicol2013,Woods2018,Woods2017,Ezawa2014}. The realization of these new effects requires free standing layers, which is challenging from an experimental point of view. Due to the $sp^2-sp^3$ orbital hybridization silicene, germanene, and stanene can interact strongly with the substrate, which may significantly affect their Dirac properties \cite{Ezawa2018}. Recent progress, however, indicates that it is possible to achieve quasi-freestanding silicene and germanene layers in the laboratory \cite{Du2016,Zandvliet2016,Tang2018,Qin2017}. 

\begin{figure*}
\includegraphics[width=\textwidth,height=5cm]{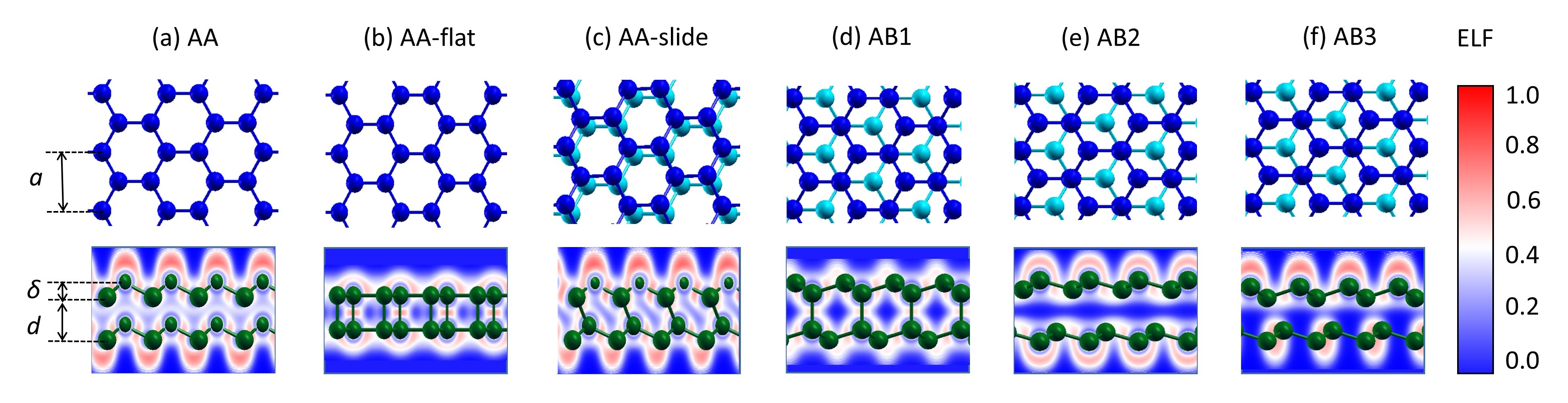}
\caption{\label{figure1} (Color online) Top and side views of the optimized stacking structures of the staggered bilayer systems. Structural parameters, such as the lattice constant $a$, staggering between the inequivalent atoms in the unit cell $\delta$, and interlayer separation $d$ are also denoted. The electron localization function, calculated for the b-Ger structures, is superimposed on the side views (bottom row). The scale for the ELF density plots is shown as a vertical bar to the right.}
\end{figure*}

In addition to hexagonal 2D materials with different atomic composition, bilayer graphene held together by van der Waals (vdW) interactions has become a separate field of research. The existence of massive Dirac carriers, Moire patterns, and unusual superconductivity are some of the key features that distinguish these weakly bonded systems from single graphene \cite{Koshino2013,Po2018}. Bilayer materials composed of staggered honeycomb layers become a natural direction of exploration with some structures already synthesized experimentally \cite{Davila2016,Nakano2016}. The stronger interaction between the layers due to pronounced $sp^3$ bonding, finite staggering, and significant SOC distinguish bilayer silicene (b-Sil), bilayer germanene (b-Ger), and bilayer stanene (b-Sta) from bilayer graphene (b-Gra). 

A limited number of computational studies have shown that the electronic structure of the staggered bilayers can be affected significantly by the mutual atomic orientation of the comprising individual layers and their stronger chemical bonding \cite{Meng2014,Kan2016}. Intrinsic magnetism and spontaneous band gap opening have also been shown for b-Sil and b-Ger \cite{Wu2017}. Additionally, it has been suggested that some b-Sil and b-Sta may exhibit properties intermediate to those of a topological insulator and a trivial band insulator \cite{Ezawa2012Japan}. Clearly the field of staggered hexagonal bilayers is still in its infancy and many questions concerning the atomic and electronic structures as well as the existence of nontrivial topology need further exploration. Specifically, the chemical interactions due to the reactive $sp^3$ bonding and the presence of weaker vdW interactions can significantly affect the stability of the different bilayers, thus these must be carefully examined. The existence of Dirac features in the electronic structure may depend on the stacking patterns and the magnitude of the SOC, and it may have different origin than the $p_z$ Dirac bands in graphene. There may be diverse topological features in which the atomic composition plays an important role. 

In this study, we utilize first principles methods to investigate in depth the effects of the compositional and structural makeup of the staggered bilayers in the graphene family. We show that there are several stacking patterns in b-Sil, b-Ger, and b-Sta, which do not occur in b-Gra due to the flatness of the graphenes. The strength of chemical interactions is also strongly dependent on the stacking patterns and the vdW interactions is found to be critical for the stability of the different systems. The electronic band structure is also affected in profound ways by the particular types of atoms and their mutual orientation. There are Dirac bands in various locations of the electronic band structures, which have a complex make-up and give rise to topological features in the Hall response of these systems. Our in-depth study shows that staggered bilayers have unique properties and they offer new directions for probing Dirac-like physics and band topology.

\section{Bilayered structures and computational framework}

We perform a series of calculations based on density functional theory (DFT) to determine the atomic, electronic, and phonon structures of b-Sil, b-Ger, and b-Sta. For this purpose, we utilize the Quantum ESPRESSO package \cite{Giannozzi2017,Giannozzi2009} within the Perdew-Burke-Ernzerhof \cite{pbe} (PBE) exchange-correlation potential.  Since the vdW interlayer interactions play an important role in these structures, we have included the computationally efficient atom pairwise DFT-D3 dispersion correction scheme \cite{dftd3}. The projector augmented-wave (PAW) method \cite{PAWmethod} and plane-wave sets with a kinetic energy cutoff of 180 Ry are used. To eliminate interactions between the periodic images a vacuum layer of 15 \AA \ is introduced between the bilayers. The Brillouin zone is sampled at $21\times 21\times 1$ uniform $k$-point grid and the bilayers are relaxed by allowing all the cell parameters to change with $10^{-6}$ eV and $10^{-4}$ eV/\AA \ energy and force relaxation criteria, respectively. Relativistic corrections due to SOC are also included in the calculations, which can be done simulteneously with the DFT-D3 vdW dispersion correction. 

\begin{table*}
\caption{\label{table1}Calculated structural and energetic parameters for the relaxed bilayers structures, where $a$ is the lattice constant, $\delta$ is the buckling, and $d$ is the interlayer distance. The difference in total energy $\Delta E=E_M - E_{\rm{AA-flat}}$ ($M=\rm{AA}, \rm{AA-slide}, \rm{AB1}, \rm{AB2}, \rm{AB3}$) with respect to the total energy $E_{\rm{AA-flat}}$ of the corresponding flat bilayers is also shown. The SOC and the vdW dispersion within DFT-D3 are included in all calculations.}
\resizebox{\textwidth}{!}{%
\begin{tabular}{|l|cccc|cccc|cccc|}
\br
\multicolumn{1}{|l|}{\multirow{2}{*}{}} & \multicolumn{4}{c|}{b-Sil} & \multicolumn{4}{c|}{b-Ger} & \multicolumn{4}{c|}{b-Sta} \\  
\multicolumn{1}{|l|}{} & $a$ (\AA) & $\delta$ (\AA) & $d$ (\AA) & $\Delta E$ (eV) & $a$ (\AA) & $\delta$ (\AA) & $d$ (\AA) & $\Delta E$ (eV) & $a$ (\AA) & $\delta$ (\AA) & $d$ (\AA) & $\Delta E$ (eV) \\ \mr
AA-flat & 4.13 & 0 & 2.41 & 0 & 4.38 & 0 & 2.54 & 0 & 5.06 & 0 & 2.93 & 0 \\
AA       & 3.76 & 0.91 & 2.74 & 0.22 & 3.99 & 1.05  & 2.93 & 0.56 & 4.59 & 1.3   & 3.32 & -0.09 \\
AA-slide & 3.81 & 0.72  & 2.85 & 0.14 & 4.02 & 1.08  & 2.96 & 0.02 & - & - & - & - \\
AB1      & 3.84 & 0.68 & 3.19 & 0.2 & 4.02 & 0.77 & 3.42 & 0.07 & 4.65 & 0.93  & 3.92 & 0.19 \\
AB2      & 3.85 & 0.52 & 3.64 & 0.7 & 4.01 & 0.72  & 3.79 & 0.48 & 4.66 & 0.88  & 4.52 & 0.61 \\
AB3      & 3.86 & 0.68 & 3 & 0.57 & 4.06 & 0.85 & 3.09 & 1.1  & 4.58 & 1.24  & 3.34 & 0.19 \\ 
\br
\end{tabular}%
}
\end{table*}

The two most recognizable orientations in b-Gra are the AA (the C hexagons from the individual layers are directly above each other) and AB (the center of one C hexagon is above a C atom from the other) stacking patterns.  Here we find that several variations of AA and AB patterns after relaxation are possible for b-Sta, b-Ger, and b-Sil. Fig. \ref{figure1}a shows the atomic structure of AA bilayers with hexagons from the individual layers positioned directly above each other such that $\delta$ for each layer follows the same pattern. The same type of AA pattern but composed of flat layers is given in Fig. \ref{figure1}b. Yet another AA variation with slightly displaced layers, but with different $\delta$ pattern for each layer is found for b-Sil and b-Ger as shown in Fig. \ref{figure1}c. There are also three distinct AB configurations for all atomic compositions, as shown in Fig. \ref{figure1}(d-f), whose differences originate from the two atoms in the unit cell and the buckling pattern of each layer. The calculated electron localization function (ELF), given as density plots in the side view perspective of each panel, can also be used to understand the interlayer interactions. The characteristic values for ELF are $0\leq ELF \leq 1$ and they reflect the degree of the electron density localization as $ELF\approx 1$ corresponds to perfect localization and $ELF\approx 0.5$ corresponds to an electron gas. The ELF results clearly indicate that the interlayer interaction is strongly dependent on the stacking pattern. Covalent-like bonds between the layers in AA-flat, AA-slide, and AB1 configurations suggest strong chemical interlayer bonding. On the other hand, the lack of such interlayer bonds in the AA, AB2, and AB3 bilayers shows that these are weakly bound, thus vdW interactions will play a more prominent role.

The results for the optimized structural parameters (Fig. \ref{figure1}a) for all bilayers are summarized in Table \ref{table1}. We find that the lattice constant increases as the atomic number increases and $a$ is the largest for the AA-flat structures within each composition. While $\delta=0$ for AA-flat, the largest buckling is calculated for the AA b-Sil and b-Sta, and AA-slide b-Ger. We also find that the interlayer distance is the smallest for AA-flat in all cases, but the farthest separation occurs for the AB2 stacking for all structures. These $d$ values correlate with the calculated ELF further indicating that the AA-flat bilayers experience the strongest chemical interaction, while the AB2 bilayers are the most weakly bound. 

The energetic stability for the different stacking patterns is also examined by calculating $\Delta E=E_M-E_{\rm{AA-flat}}$. Here, $E_M$ is the total energy for $M=\rm{AA}, \rm{AA-slide}, \rm{AB1}, \rm{AB2}, \rm{AB3}$ and the total energy for the AA-flat, $E_{\rm{AA-flat}}$, is taken as a reference point. It is found that in all cases the AA-flat is the most energetically stable with the exception of b-Sta for which the AA is preferred over a small difference $\Delta E = 0.09$ eV. The AA-slide configurations have a comparable to the AA total energy, with $\Delta E = 0.14$ eV for b-Sil and $\Delta E = 0.02$ eV for b-Ger, followed by the AB1 stacks, which are the most favorable among the AB structures. It was also obtained that the AA-slide b-Sta is not stable after structural relaxation. 

The phonon band structure for all bilayers is also computed in the quasi-harmonic aproximation by using the PHonon code \cite{Giannozzi2017} combined with the density functional perturbation theory (DFPT) method in Quantum ESPRESSO, and the results are shown in Figs. S1-S3 in the Supplementary information. The lack of imaginary frequencies implies that all the structures are dynamically stable.

\begin{figure*}
\includegraphics[width=\textwidth,height=9.5cm]{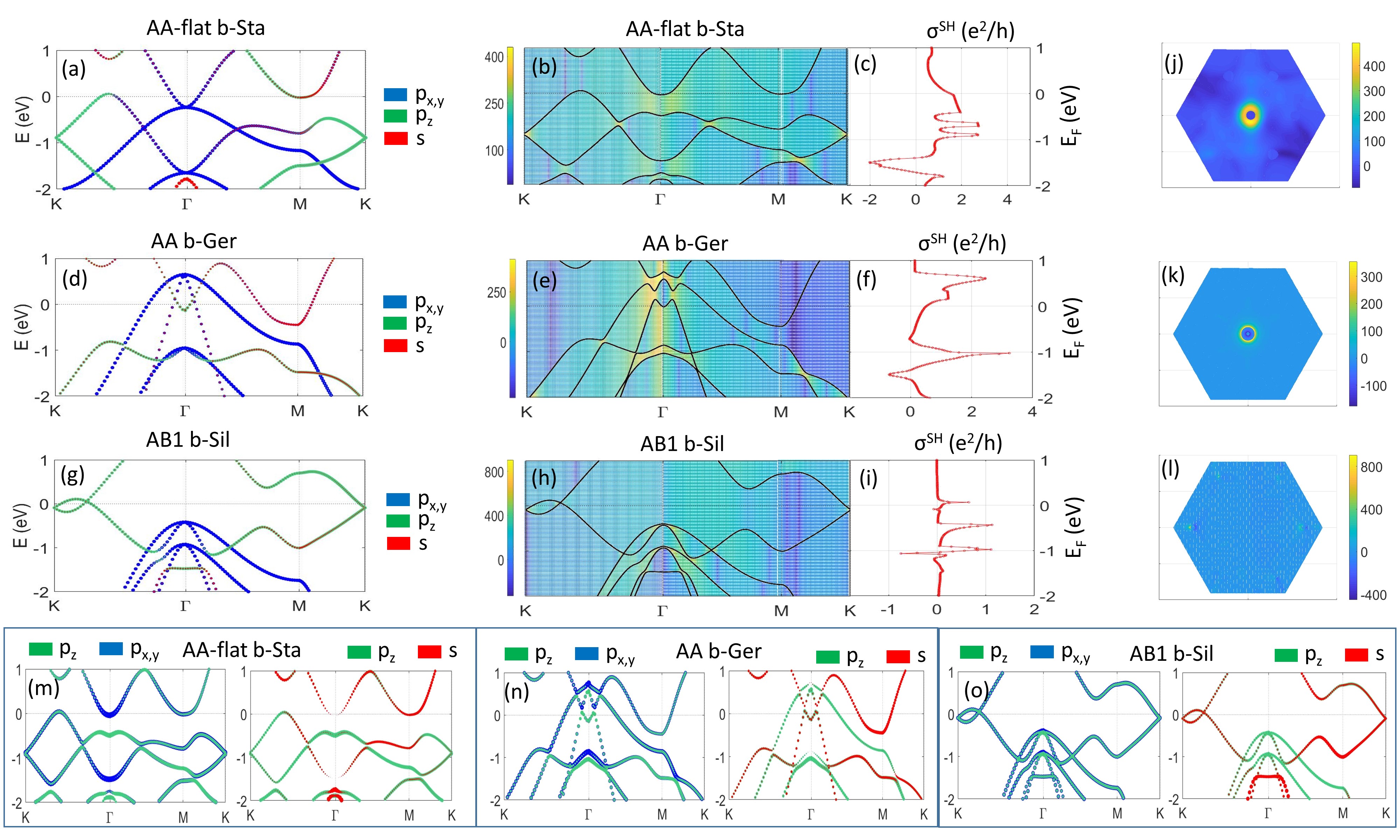}
\caption{\label{figure2} (Color online) Orbital-projected electronic band structures with no SOC included for (a) AA-flat b-Sta, (d) AA b-Ger, and (g) AB1 b-Sil. Electronic band structure with SOC included for (b) AA-flat b-Sta, (e) AA b-Ger, and (h) AB1 b-Sil. The color intensity map inside each panel depicts the spin Berry curvature, calculated along the same k-space symmetry lines as $\sum_n \Omega_{n, xy}^{(SH)}$ with $\Omega_{n, xy}^{(SH)}$ given by Eq. \ref{omega_sh}. The intrinsic spin Hall conductivity (Eq. \ref{sigma}) as a function of the chemical potential position is shown for (c) AA-flat b-Sta, (f) AA b-Ger, and (i) AB1 b-Sil. Density plots for the spin Berry curvature in reciprocal space at the Fermi energy are given for (j) AA-flat b-Sta, (k) AA b-Ger, and (l) AB1 b-Sil. Orbital-projected electronic band structures with SOC included are shown for (m) AA-flat b-Sta, (n) AA b-Ger, and (o) AB1 b-Sil, respectively. All calculations are performed with DFT-D3 vdW interaction.}
\end{figure*}

\section{Electronic Structure and Topology}

The diverse AA and AB stacking patterns affect the electronic structures of the bilayers in distinct ways for each composition. To illustrate this diversity, in Fig. \ref{figure2} we present the particular cases of AA-flat b-Sta, AA b-Ger, and AB1 b-Sil. Fig. \ref{figure2} (a, d, g) show the corresponding band structures with explict orbital projection (discussed below) and with no SOC. These results indicate that bands at the $K, \Gamma, $ and $M$ points have different characteristics. For example, there is a Dirac band crossing at the K-point well below the Fermi level (at $E=-1$ eV) for the AA-flat b-Sta, there are two Dirac band crossings near the Fermi level at and around the K-point for the AB1 b-Sil, while no band crossing at the K-point is found for the AA b-Ger. The band behavior at the $\Gamma$ point is also distinct for the considered cases. There are two touching bands slightly under the Fermi level for the AA-flat b-Sta, however these bands move up in the valence range with two emerging crossing points around the $\Gamma$ point for the AA b-Ger. Also, there is a large gap for the AB1 b-Sil at $\Gamma$, reminiscent of the band structure of a single layer silicene \cite{silicenereview}.

The SOC plays a distinct role for each bilayer, and its role increases with the atomic number of each composition. To illustrate this, Figs. \ref{figure2} (b, e, h) show the electronic band structure for the same cases calculated with the inclusion of the SOC. There are several band gaps opened in various locations of the band structures. For example, the SOC gap is 41 meV at the K-point, and 0.378 eV to the left and right of the $\Gamma$-point for AA-flat b-Sta (Fig. \ref{figure2}b). The gaps around the $\Gamma$ point for AA b-Ger are 97 meV (Fig. \ref{figure2}e), while for AB1 b-Sil they are much smaller, about 30 meV and 25 meV at the $\Gamma$ point and close to the K-point, respectively (Fig. \ref{figure2}h). 

Let us further note that the vdW interaction also induces changes in both the atomic and electronic structures of the bilayers. Typically the buckling parameter within DFT-D3 is larger within 0.006 \AA \ - 0.02 \AA \ range for all structures when compared with results without the vdW correction, with  the smallest effect found for AA-flat b-Sil and the largest for AA b-Sta. The vdW correction also decreases the interlayer distance with the largest reduction of 0.05 \AA \ for AA b-Sil structure when compared to results with standard DFT calculations. 

These structural changes also affect the energy bands of some of the bilayers. For the chemically bonded bilayers, such as the AA-flat, AA-slide and AB1, the band structures are not changed significantly. Specifically, the regions with the band crossings at the Fermi level near the K-point, characteristic of the AB1 stacked structures, remain practically unaffected.  For the weakly bonded layers, however, some band shifts are observed. For example, the vdW inclusion in AA b-Ger results in a down shift of the conduction bands at the $\Gamma$-point by about 0.1 eV when compared to results from standard DFT calculations (Fig. S4(a) in the Supplementary Information). At the same time, this weak interaction leads to a shift to higher energies of the valence bands at the $\Gamma$-point of about 0.15 eV for b-Sta (Fig. S4 (b) in the Supplementary Information).

To get a better insight into the SOC induced band gaps and its effects on the hybridization at different location of the band structure, we also determine the orbital band composition for the different bilayers by showing the $p_{x,y,z}$ and $s$ contributions in a color coded scheme while distinguishing between cases with no SOC (Fig. \ref{figure2}(a, d, g)) and with SOC included in the calculations (Fig. \ref{figure2}(m, n, o)). One finds that without SOC the makeup of the bands near the K-point is predominantly of $p_z$ character fo the AA-flat stacking (Fig. \ref{figure2}(a, g)). The SOC, however, results in admixture with the $p_{x,y}$ states for the AA-flat (Fig. \ref{figure2}(m)) and admixture with the $p_{x,y}$ and $s$ states for the AB1 (Fig. \ref{figure2}(o)) bilayers. Since the $p_{x,y}$ states are part of the $\sigma$ bond, these complex Dirac states are, unlike the $p_z$ Dirac orbitals, quite robust against deformations or substrate interactions \cite{Oshiyama2013}.

The makeup of the bands near the $\Gamma$ point is also complex. Regardless of the SOC presence there is an admixture of  $p_{x,y,z}$ and $s$ states near $\Gamma$, however this relativistic correction induces gaps which may lead to changes in the orbital character. For example, for the case of b-Sta (Fig. \ref{figure2}m) at the $\Gamma$-point, there is a band inversion between the $s$ and $p_{x,y}$ states at about -1.8 eV below the Fermi level, and a band inversion between the $p_{x,y}$ and $p_z$ states immediately below the Fermi level, rendering a non-trivial topology of the band structure. Similarly, for AA b-Ger at the $\Gamma$-point there are band invesrion changes in orbital character near the Fermi level between the $p_{x,y}$ and $p_z$ states (Fig. \ref{figure2}(n) left), and between the $p_{z}$ and $s$ states (Fig. \ref{figure2}(n) right).

\begin{figure*}
\includegraphics[width=\textwidth,height=6cm]{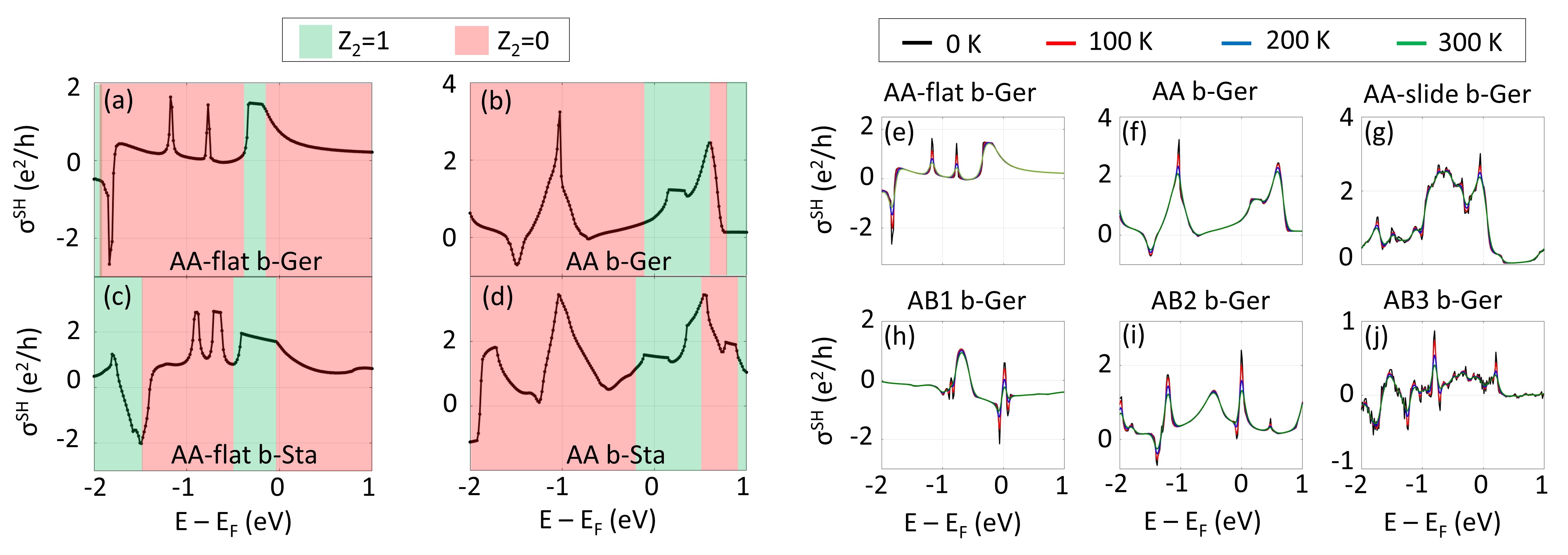}
\caption{\label{figure3} (Color online) The intrinsic spin Hall conductivity dependence on the chemical potential for (a) AA-flat b-Ger, (b) AA b-Ger, (c) AA-flat b-Sta, and (d) AA b-Sta structures. The green and red background shades correspond to regions with topological index $\rm{Z}_2=1$ and $\rm{Z}_2=0$, respectively. The the spin Hall conductivity as a function of the chemical potential is also shown for all b-Ger configurations in (e)-(l) for different values of the temperature. All data is shown in $e^2/h$ units.}
\end{figure*}

The full compositional and stacking variety for all orbitally projected band structures with and without SOC is on display in Figs. S5, S6, and S7 in the Supplementary Information. One notes that band inversion occurs only for select bilayers, such as the AA-flat b-Sta and b-Ger, or AA b-Sta and b-Ger (as discussed for Fig. \ref{figure2}(m,n)). All other bilayers do not exhibit such changes due to SOC inclusion in their band structures. 
 
The topological character of the different band structures is further investigated by considering the intrinsic Hall conductivity response. Given that these 2D materials have time reversal symmetry, their intrinsic anomalous Hall conductivity is zero. At the same time, the strong SOC in these metallic systems indicates that these bilayers have a nonzero intrinsic spin Hall conductivity. Here this response property is computed within the Kubo formalism \cite{Sinova2015}:
\begin{equation}
\sigma^{(SH)}_{xy} =  e \sum_{n} \int_{BZ} \frac{d \mathbf{k}}{(2\pi)^2} f_{n \mathbf{k}} \Omega_{n}^{(SH)} (\mathbf{k}) \label{sigma}
\end{equation}
 \begin{equation}
\Omega_{n, xy}^{(SH)} = -2 \rm{Im} \sum_{m \neq n} \frac{\langle u_{n \mathbf{k}} | \hat{J}_x^z | u_{m \mathbf{k}} \rangle  \langle u_{m \mathbf{k}} | \hat{v}_y | u_{n \mathbf{k}} \rangle}{(\epsilon_{n \mathbf{k}} - \epsilon_{m \mathbf{k}})^2} \label{omega_sh}
\end{equation}
where $\Omega_{n, xy}^{(SH)}$ is the spin Berry curvature of the $n$-th band, and $\epsilon_{n \mathbf{k}}$ and $| u_{n \mathbf{k}} \rangle$ are the eigenvalues and eigenvectors of each system's Bloch Hamiltonian $\hat{H}$. The spin current operator $\hat{J}_x^z = \{ \hat{v}_x , \hat{\sigma}_z \} /2$ describes the spin flow with $z$-polarization along the $x$-direction given by the velocity operator $\hat{v}_x= \hbar^{-1} d \hat{H} / d k_x$ and the Pauli spin operator $\hat{\sigma}_z$. Also, 
$f_{n \mathbf{k}}= 1/(1+e^{ (\epsilon_{n \mathbf{k}} - E_{F}) / k_B T})$ is the Fermi-Dirac distribution. In order to make the integration over the Brillouin zone in Eq. \ref{sigma} computationally efficient, we employ a Wannier interpolation technique \cite{Vanderbilt2006}, where the \it{ab initio} \rm wave functions for each phase are projected on 8 maximally-localized Wannier functions per atom, corresponding to the spinfull $s$ and $p$ orbitals, with the Wannier90 code \cite{wannier}.

The color intensity maps, superimposed on the band structures in Fig. \ref{figure2} (b, e, h), give a visual representation of the spin Berry curvature as a function of both the momentum $k$ and the position of the Fermi level $E_F$ for AA-flat b-Sta, AA b-Ger and AB1 b-Sil, respectively. The figure shows that the SOC induced band gaps are indeed sources for the spin Berry curvature. While the magnitude of $\Omega^{(SH)}$ is always increased in the vicinity of avoided band crossings, it is particularly enhanced around regions with band inversion, such as $E$=-1.8 eV for AA-flat b-Sta, $E \approx$ 0 eV for AA b-Ger near the $\Gamma$-point, $E$=-0.2 eV for AA-flat b-Sta at the $\Gamma$-point, or $E \approx$ -1 eV for AA b-Ger along the $\Gamma$-K line.

A different perspective of $\Omega^{(SH)}$ is shown in Fig. \ref{figure2}(j, k, l) displaying the distribution of the spin Berry curvature at the Fermi level over the reciprocal space. One notices the rings around the $\Gamma$ point for the case of AA-flat b-Sta and AA b-Ger, and the peaks on the $\Gamma$-K line for AB1 b-Sil, which correspond to the band inversion regions of the bilayers. 

The numerical results for the intrinsic spin Hall conductivity as a function of chemical potential are shown in Fig. \ref{figure3}. Of particular interest are the nearly quantized plateaus for the AA-flat and AA b-Sta and b-Ger in Fig. \ref{figure3} (a-d). These are associated with the chemical potential residing in SOC induced gaps, which also have band inversion. The behavior of these quantized peaks can be qualitatively understood in the context of a two-band gapped Dirac model in 2D. In such a simplified approach, the spin Hall conductivity can be obtained analytically \cite{Woods2017,Woods2018} with $\sigma_{xy}^{(SH)}(E_F) = \frac{e}{2} \sum_{i} \frac{\Delta_i}{\rm{Max}\left[ \abs{\Delta_i}, \abs{E_F} \right]}$, where  $\Delta_i$ is the  Dirac mass gap for the $i$-th cone. Clearly, $\sigma_{xy}^{(SH)}(E_F)$ is quantized when $E_F$ resides in the region of the gap and the quantization value (in units of $e^2/h$) is determined by the number of cones. As $E_F$ is outside of the gap for a given cone, the spin Hall conductivity falls off as $1/E_F$. The devitations from this qualitative behavior in Fig. \ref{figure3} are attributed to the more complicated band structure of the bilayers coming from other bands crossing the Fermi surface at different $k$-regions in the Brillouin zone. Other effects, such as the contribution from all occupied bands in the Kubo formalism and the nontrivial $k$-dependence of the bands in the vicinity of the SOC gaps are also important when analyzing the nearly quantized regions in the spin Hall conductivity of the bilayers.

The topological nature of electronic bands for materials with inversion symmetry can also be classified by using the Z2 index, whose values of 0 and 1 distinguish between trivial and nontrivial band topology, respectively. This index takes into account the parity of occupied bands at time-reversal-invariant-momenta points (TRIM) in the reciprocal space\cite{Kane2007}, according to Z2=$(-1)^{\nu}=\prod_i \prod_m \xi_m (\Lambda_i)$, where $\xi_m$ are the parities of the occupied Kramer's doublets at TRIM $\Lambda_i$. Given that the considered bilayers in Fig. \ref{figure3}(a-e) have time-reversal and inversion symmetries, the Z2 index captures the nontrivial topology despite that these systems are not globally gapped (other bands cross $E_F$ at different $k$ while $E_F$ resides in the SOC gap around the $\Gamma$ point, for example). The calculated values for Z2 are shown in Fig. \ref{figure3}(a-e) with green and red background colors, corresponding to the regions of non-trivial (Z2=1) or trivial (Z2=0) band topologies, respectively. We notice that the regions with Z2=1 coincide with the plateaus of the spin Hall conductivity, indicating that the SOC gaps with band inversion induce a non-trivial topology in these particular structures. We mention that the AB3 and AA-slide bilayer configurations do not preserve the inversion symmetry, and therefore for these cases only the spin Hall conductivity gives an appropriate description of their electronic band topological nature.

We also investigate the temperature influence on the spin Hall response, as controlled by the Fermi distribution function in Eq. \ref{sigma}. The results for all the b-Ger structures are shown in Fig. \ref{figure3} (e-j). One sees that, while generally the effects of $T$ are to decrease the absolute value of spin Hall conductivity and to smooth out its $E_F$ dependence, the plateau regions of $\sigma_{xy}^{(SH)}$ are quite robust against changes in $T$. The numerical results for the $\sigma_{xy}^{(SH)}$ dependence on the chemical potential for all the b-Sil and b-Sta structures are shown in Fig. S8 and S9 in the Supplementary Information.

\section{Conclusions}

In summary, we investigate from first principles the compositional and stacking dependent poperties of bilayered materials from the extended graphene family. Unlike the flat graphenes, the staggered structure of silicene, germanene, and stanene allow several variations of the AA and AB configurations with distinct features in the electronic band structure. The structural stability is influenced by the $sp^2-sp^3$ hybridization and dispersion interactions, which dominate different bilayers in a distinct manner, resulting in chemically bonded or weakly bonded systems. It is quite interesting to note that the interplay between the atomic number, SOC magnitude and stacking patterns renders a non-trivial band topology in select AA and AA-flat arrangements. 

The metallic nature and the presence of SOC ensures that these bilayers have an intrinsic spin Hall conductivity, which is calculated using the Kubo approach, and its dependence on the chemical potential and temperature is related to the distribution of the spin Berry curvature in reciprocal space. Selected bilayers, however, can be characterized as quantum spin Hall insulators due to their (nearly) quantized spin Hall conductivities, which are also fairly robust as a function of temperature. The quantized nature of the spin Hall response for the select bilayers is attributed to a complex band inversion in SOC band gaps. In addition to this, the interpretation in the context of the Z2 topological number suggests the appearance of protected by time reversal edge states in AA and AA-flat b-Ger and b-Sta. 

The experimental realization of such 2D bilayered systems depends on appropriate substrates.  Several studies have shown that in many cases the chemical bonds with the substrate can significantly affect the Dirac-like properties of silicene, germanene, and stanene as summarized in \cite{Molle2017}. Recent reports, however, indicate that weakly bonded staggered hexagonal layers with preserved Dirac states may be possible on SiC and MoS2 substrates \cite{MacDonald2017,Zand2016}, which brings optimism that such substrates maybe useful for the studied bilayers here. Our study adds another direction for basic science properties in the area of graphene-like materials. As progress in the synthesis of these systems is made in the laboratory, bilayered silicene, germanene, or stanene can become a new playground for topology tuning via stacking and compositional routes.

\ack{
Financial support from the US Department of Energy under Grant No. DE-FG02-06ER46297 is acknowledged. P.R.-L. also acknowledges partial support from TerMic (Grant No. FIS2014-52486-R, Spanish Government), CONTRACT (Grant No. FIS2017-83709-R, Spanish Government) and from Juan de la Cierva - Incorporacion program (Ref: I JCI-2015-25315, Spanish Government). 
}


\section*{References}

\bibliography{bilayers}

\providecommand{\newblock}{}
\begin{thebibliography}{10}
\expandafter\ifx\csname url\endcsname\relax
  \def\url#1{{\tt #1}}\fi
\expandafter\ifx\csname urlprefix\endcsname\relax\def\urlprefix{URL }\fi
\providecommand{\eprint}[2][]{\url{#2}}

\bibitem{LeFay2012}
Vogt P, De~Padova P, Quaresima C, Avila J, Frantzeskakis E, Asensio M~C, Resta
  A, Ealet B and Le~Lay G 2012 {\em Phys. Rev. Lett.\/} {\bf 108} 155501

\bibitem{Takamura2012}
Fleurence A, Friedlein R, Ozaki T, Kawai H, Wang Y and Yamada-Takamura Y 2012
  {\em Phys. Rev. Lett.\/} {\bf 108} 245501

\bibitem{Kawai2012}
Lin C~L, Arafune R, Kawahara K, Tsukahara N, Minamitani E, Kim Y, Takagi N and
  Kawai M 2012 {\em Appl. Phys. Express\/} {\bf 5} 045802

\bibitem{silicenereview}
Chowdhury S and Jana D 2016 {\em Rep. Prog. Phys.\/} {\bf 79} 126501

\bibitem{LeLay2014}
Dávila M~E, Xian L, Cahangirov S, Rubio A and Lay G~L 2014 {\em New J.
  Phys.\/} {\bf 16} 095002

\bibitem{Pirri2015}
Derivaz M, Dentel D, Stephan R, Hanf M~C, Mehdaoui A, Sonnet P and Pirri C 2015
  {\em Nano Lett.\/} {\bf 15} 2510

\bibitem{Jia2015}
feng Zhu F, jiong Chen W, Xu Y, lei Gao C, dan Guan D, hua Liu C, Qian D, Zhang
  S~C and feng Jia J 2015 {\em Nat. Mater.\/} {\bf 14} 1020

\bibitem{germanenereview}
Acun A and et~al 2015 {\em J. Phys.: Condens. Matter\/} {\bf 27} 443002

\bibitem{Shukla2016}
Saxena S, Chaudhary R~P and Shukla S 2016 {\em Scientific Reports\/} {\bf 6}
  31073

\bibitem{Ciraci2009}
Cahangirov S, Topsakal M, Akt\"urk E, \ifmmode~\mbox{\c{S}}\else \c{S}\fi{}ahin
  H and Ciraci S 2009 {\em Phys. Rev. Lett.\/} {\bf 102} 236804

\bibitem{Bechstedt2013}
Matthes L, Pulci O and Bechstedt F 2013 {\em J. Phys.: Condensed Matter\/} {\bf
  25} 395305

\bibitem{Smallreview2015}
Balendhran S, Walia S, Nili H, Sriram S and Bhaskaran M 2015 {\em Small\/} {\bf
  6} 640

\bibitem{Ezawa2012}
Ezawa M 2012 {\em Phys. Rev. Lett.\/} {\bf 109} 055502

\bibitem{Ezawa2013}
Ezawa M 2013 {\em Phys. Rev. Lett.\/} {\bf 110} 026603

\bibitem{Nicol2013}
Tabert C~J and Nicol E~J 2013 {\em Phys. Rev. Lett.\/} {\bf 110} 197402

\bibitem{Woods2018}
Rodriguez-Lopez P, Kort-Kamp W~J~M, Dalvit D~A~R and Woods L~M 2018 {\em Phys.
  Rev. Materials\/} {\bf 2} 014003

\bibitem{Woods2017}
Rodriguez-Lopez P, Kort-Kamp W~J, Dalvit D~A and Woods L~M 2017 {\em Nat.
  Commun.\/} {\bf 8} 14699

\bibitem{Ezawa2014}
Ezawa M 2014 {\em J. Supercond. Nov. Magn.\/} {\bf 28} 1249

\bibitem{Ezawa2018}
Ezawa M, Salomon E, Padova P~D, Solonenko D, Vogt P, Davila M~E, Molle A, Angot
  T and Lay G~L 2018 {\em Rivista del Nuovo Cimento\/} {\bf 41} 3

\bibitem{Du2016}
Du Y, Zhuang J, Wang J, Li Z, Liu H, Zhao J, Xu X, Feng H, Chen L, Wu K, Wang X
  and Dou S~X 2016 {\em Sci. Adv.\/} {\bf 2} 7

\bibitem{Zandvliet2016}
Zhang L, Bampoulis P, Rudenko A~N, Yao Q, van Houselt A, Poelsema B, Katsnelson
  M~I and Zandvliet H~J~W 2016 {\em Phys. Rev. Lett.\/} {\bf 116} 256804

\bibitem{Tang2018}
Lin C~H, Huang A, Pai W~W, Chen W~C, Chen T~Y, Chang T~R, Yukawa R, Cheng C~M,
  Mou C~Y, Matsuda I, Chiang T~C, Jeng H~T and Tang S~J 2018 {\em Phys. Rev.
  Materials\/} {\bf 2} 024003

\bibitem{Qin2017}
Qin Z, Pan J, Lu S, Shao Y, Wang Y, Du S and Gao H 2017 {\em Adv. Mater.\/}
  {\bf 29} 13

\bibitem{Koshino2013}
McCann E and Koshino M 2013 {\em Rep. Prog. Phys.\/} {\bf 76} 056503

\bibitem{Po2018}
{Po} H~C, {Zou} L, {Vishwanath} A and {Senthil} T 2018  (\textit{Preprint}
  \eprint{arXiv:1803.09742})

\bibitem{Davila2016}
Dávila M~E and Lay G~L 2016 {\em Scientific. Reports\/} {\bf 6} 20714

\bibitem{Nakano2016}
Yaokawa R, Ohsuna T, Morishita T, Hayasaka Y, Spencer M~J and Nakano H 2016
  {\em Nat. Commun.\/} {\bf 7} 10657

\bibitem{Meng2014}
Fu H, Zhang J, Ding Z, Li H and Meng S 2014 {\em Appl. Phys. Lett.\/} {\bf 104}
  131904

\bibitem{Kan2016}
Huang C, Zhou J, Wu H, Deng K, Jena P and Kan E 2016 {\em J. Phys. Chem.
  Lett.\/} {\bf 7} 1919

\bibitem{Wu2017}
Wang X and Wu Z 2017 {\em Phys. Chem. Chem. Phys.\/} {\bf 19} 2148

\bibitem{Ezawa2012Japan}
{Ezawa} M 2012 {\em J. Phy. Soc. Jpn.\/} {\bf 81} 104713

\bibitem{Giannozzi2017}
Giannozzi P and et~al 2017 {\em J. Phys. Condens. Matter\/} {\bf 29} 465901

\bibitem{Giannozzi2009}
Giannozzi P and et~al 2009 {\em J. Phys. Condens. Matter\/} {\bf 21} 395502

\bibitem{pbe}
Perdew J~P, Burke K and Ernzerhof M 1996 {\em Phys. Rev. Lett.\/} {\bf 77} 3865

\bibitem{dftd3}
Grimme S, Antony J, Erlich S and Krieg H 2010 {\em J. Chem. Phys.\/} {\bf 132}
  154104

\bibitem{PAWmethod}
Bl\"ochl P~E 1994 {\em Phys. Rev. B\/} {\bf 50} 17953

\bibitem{Oshiyama2013}
Guo Z~X, Furuya S, Iwata J~i and Oshiyama A 2013 {\em Phys. Rev. B\/} {\bf 87}
  235435

\bibitem{Sinova2015}
Sinova J, Valenzuela S~O, Wuderlich J, Back C~H and Jungwirth T 2015 {\em Rev.
  Mod. Phys.\/} {\bf 87} 1213

\bibitem{Vanderbilt2006}
Wang X, Yates J~R, Souza I and Vanderbilt D 2006 {\em Phys. Rev. B\/} {\bf 74}
  195118

\bibitem{wannier}
Mostofi A~A, Yates J~R, Lee Y~S, Souza I, Vanderbilt D and Marzari N 2008 {\em
  Comput. Phys. Comm.\/} {\bf 178} 685

\bibitem{Kane2007}
Fu L and Kane C~L 2007 {\em Phys. Rev. B\/} {\bf 76} 045302

\bibitem{Molle2017}
Molle A, Goldberger J, Houssa M, Xu Y, Zhang S~C and Akinwande D 2017 {\em
  Nature Mater.\/} {\bf 16} 163

\bibitem{MacDonald2017}
Li P, Li X, Zhao W, Chen H, Chen M~X, Guo Z~X, Feng J, Gong X~G and MacDonald
  A~H 2017 {\em Nano Lett.\/} {\bf 17} 6195--6202

\bibitem{Zand2016}
Zhang L, Bampoulis P, Rudenko A~N, Yao Q, van Houselt A, Poelsema B, Katsnelson
  M~I and Zandvliet H~J~W 2016 {\em Phys. Rev. Lett.\/} {\bf 116} 256804

\end{thebibliography}
\bibliographystyle{iopart-num}

\end{document}